\documentclass{article}
\usepackage{fancyhdr}
\usepackage{graphicx}					% including graphic
\usepackage{amsmath, amsthm, amssymb}	%math symbols
\usepackage{icomma}						% for correct deciaml separator

\usepackage[utf8]{inputenc} 
\usepackage[T1]{fontenc} 
 
\usepackage{tabularx}
\newcolumntype{L}[1]{>{\raggedright\arraybackslash}p{#1}} % linksbündig mit Breitenangabe
\newcolumntype{C}[1]{>{\centering\arraybackslash}p{#1}} % zentriert mit Breitenangabe
\newcolumntype{R}[1]{>{\raggedleft\arraybackslash}p{#1}} 
\usepackage{booktabs}

\usepackage{subcaption}  % To place figures side by side, if you wish to do that in latex... (package "subfigure" is deprecated)

\begin{document}

\LARGE{\bf Nearest-Neighbor Based Non-Parametric Probabilistic Forecasting with Applications in Photovoltaic Systems}
\Large 

Jorge \'Angel Gonz\'alez Ordiano, Wolfgang Doneit, Simon Waczowicz, Lutz Gr\"oll, Ralf Mikut, Veit Hagenmeyer\\
\normalsize
{\it Karlsruhe Institute of Technology, Institute for Applied Computer Science\\}
{\it Hermann-von-Helmholtz-Platz 1, 76344 Eggenstein-Leopoldshafen}\\
{\it EMail: jorge.ordiano@kit.edu }

%\pagestyle{empty}
%\tableofcontents

\setcounter{page}{1}

\section{Introduction}

Time series forecasting (i.e. the prediction of unknown future time series values using known data) has found several applications in a number of fields, like, economics and electricity forecasting~\cite{Maimon05}. Most of the used forecasting models deliver a so-called point forecast~\cite{Chatfield93}, a value that according to the models' criteria is most likely to occur. Nonetheless, such forecasts lack information regarding their uncertainty. A possibility of quantifying such uncertainty is by conducting probabilistic forecasts~\cite{Gneiting14,Hong16}, which can be delivered as prediction intervals (including the probability of the forecast being inside the interval) or complete conditional probability distributions of future time series values~\cite{Juban07}. Such a quantification of the forecast uncertainty is of interest for several optimization problems, as e.g. model predictive control. Probabilistic forecasting is divided in parametric and non-parametric approaches. While the former assume that the forecast values follow a known distribution (e.g. Gaussian) and try to determine the parameters describing it, the latter make no assumptions, but instead attempt to approximate the underlying distribution via the training data. Non-parametric approaches have the advantage of not assuming that all values will follow the same probability distribution across all points in time~\cite{Charytoniuk99,Zhang14}. 

Quantile regression is an approach that has been utilized in several works (e.g. Bremnes~\cite{Bremnes04} and Nielsen et al.~\cite{Nielsen06}) in order to create non-parametric probabilistic forecasts. It is a type of regression that provides the conditional quantile of an output value given a certain input, and it is obtained through the minimization of a non-differentiable loss-function~\cite{Fahrmeir09}. By combining several of such quantile regression models (i.e. point forecasting models), probabilistic forecasts can be conducted~\cite{Ma15b}. However, the creation and implementation of quantile regressions, especially with more complex data-driven regression techniques, as artificial neural networks, is not a simple task, due to its non-differentiable loss function~\cite{Cannon11}. For this reason, the present contribution offers a simple method of obtaining quantile regressions, which can be combined in data-driven probabilistic forecasts, by changing the models' training set using a k-nearest-neighbor approach, traditional regression methods, and the assumption of a dense dataset.

The potential of the presented methodology for the obtainment of probabilistic forecasting models is demonstrated for the usecase of photovoltaic (PV) power forecasting.  The periodicity and length of the utilized PV power time series helps to assure the dataset's density that is necessary for the present contribution's method. Accurate PV forecasting models are of major importance, due to the PV systems' dependence on the weather that makes their electrical power generation uncertain. Such uncertainty complicates the balancing of demand and supply in the electrical grid~\cite{Waczowicz16}.

The present contribution is divided as follows: First, the principles of forecasting models are described. Then, the general framework of the data-driven non-parametric nearest-neighbor based probabilistic forecasting approach is presented. Afterwards, the principles behind the present work's photovoltaic power forecasting approach as well as the utilized dataset are shown. The next section shows and discusses the obtained results. Finally, a conclusion and an outlook are offered.

\section{Forecasting Models}

Data-driven point forecasting models are used to approximate the values of a desired time series $y$ (whose discrete time steps are denoted by $k \in [1,K]$) at a forecast horizon $H$ (with $H\geq1$) using its current or past values and/or values of other exogenous time series, $\mathbf{u}$. For example, the functional relation given by a linear or non-linear autoregressive exogenous (ARX or NARX) forecasting model that utilizes the current and past information from timestep $k$ to timestep $k-H_{1}$ of its input time series and whose parameters are represented by a vector $\boldsymbol{\theta}$, can be written as\footnote{The functional relation in Equation~(\ref{eq. point_forecast_functional_relation}) can also describe autoregressive (AR) models, if the exogenous time series $\mathbf{u}$ are not used as input values.}
\begin{equation}\label{eq. point_forecast_functional_relation}
\hat{y}[k+H] = f(y[k], ..,y[k-H_{1}],\mathbf{u}^{T}[k],..,\mathbf{u}^{T}[k-H_{1}];\boldsymbol{\theta}); k > H_{1} \text{ .}
\end{equation}
%with the vector $\boldsymbol{\theta}$ containing the parameters defining the model.

The obtainment of data-driven point forecasting models can be generalized as a regression problem, i.e. the models are obtained through a learning process in which the value of a model's loss function in a training set is minimized. For example, models obtained by minimizing the sum of squared errors (as is the case with typical data mining techniques, like polynomial regression or artificial neural networks) deliver as forecast an approximation of the conditional expected value given their used input~\cite{Hastie08,Juban07}. 

A probabilistic forecast can be attained by combining several point forecasting models which instead of delivering the conditional expected value, deliver an approximation of a conditional $q$-quantile, $\hat{y}_{q}[k+H]$ (with $q\in[0.01;0.99]$). The functional relation of such models, using the same input as in Equation~(\ref{eq. point_forecast_functional_relation}), is written in the present contribution as
\begin{equation}\label{eq. quantile_forecast_functional_relation}
\hat{y}_{q}[k+H] = f_{q}(y[k], ..,y[k-H_{1}],\mathbf{u}^{T}[k],..,\mathbf{u}^{T}[k-H_{1}];\boldsymbol{\theta}_{q}) \text{ .}
\end{equation}
Just as in Equation~(\ref{eq. point_forecast_functional_relation}), the vector $\boldsymbol{\theta}_{q}$ represents the parameters defining the model. Equation~(\ref{eq. quantile_forecast_functional_relation}) describes a quantile regression and it is obtained through the minimization of the sum of the so-called pinball-loss~\cite{Bremnes04,Fahrmeir09}, which is given as
\begin{equation}\label{eq: pinball_loss}
 L_{q} = \sum_{k=H_{1}+1}^{K-H} \begin{cases} (q-1)~(y[k+H]-\hat{y}_{q}[k+H]) & \text{ if } y[k+H] < \hat{y}_{q}[k+H] \\
			                              q~(y[k+H]-\hat{y}_{q}[k+H])    & \text{ if } y[k+H] \geq \hat{y}_{q}[k+H] \\
                                        \end{cases} \text{ .} 
\end{equation} 
Unfortunately, the pinball-loss is non-differentiable, that makes its minimization with standard algorithms and data mining techniques a complicated task. For such reason, the present contribution offers an alternative that allows the creation of quantile regression models using traditional approaches. %the traditional minimization of the sum of square errors

\section{Method}

The present section describes the principles behind the developed k-nearest-neighbor based non-parametric probabilistic forecasting approach. Those principles are divided into how the required data-driven quantile regressions are obtained and how those are combined forming intervals which can be used as probabilistic forecasting models. Additionally, the evaluation values, utilized in the present contribution to determine the accuracy of the created models, are presented. As previously stated, the creation of point forecasting models (i.e. the quantile regressions) can be generalized into a regression problem, hence, the methodology description is formulated in a general manner.

\subsection{k-Nearest-Neighbor Based Quantile Regression}

In order to solve a regression problem a training set containing a series of learning pairs is required. The training set is comprised of $N$  desired output values, $y_{n}$, and their corresponding input vectors $\mathbf{x}_{n}^{T} = (x_{n1},..,x_{nS})$, with $S$ being the number of its contained features. The learning pairs are normally written as an input matrix $\mathbf{X}$ of dimensions $N \times S$ and a $N$ dimensional desired output vector $\mathbf{y}$:
\begin{equation}\label{eq:input_matrix}
\mathbf{X} =  \begin{pmatrix}
				\mathbf{x}_{1}^{T}\\
				\vdots\\
				\mathbf{x}_{N}^{T} 
 			\end{pmatrix}  ,~
 			\mathbf{y} =  \begin{pmatrix}
				y_{1}\\
				\vdots\\
				y_{N}
 			\end{pmatrix} \text{ .}
\end{equation} 
Both $\mathbf{x}_{n}^{T}$ and $y_{n}$ can be defined for the time series case shown in Equation~(\ref{eq. point_forecast_functional_relation}) as 
\begin{equation}
\begin{aligned}
& \mathbf{x}_{n}^{T} = (y[n+H_{1}], ..,y[n],\mathbf{u}^{T}[n+H_{1}],..,\mathbf{u}^{T}[n]) \text{ ,} \\
& y_{n} = y[n+H+H_{1}] \text{ , for }\\
& n \in [1,..,N] \text{ , with } N = K-H-H_{1} \text{ .}
\end{aligned}
\end{equation}
The approach in the present contribution assumes that each input vector has several similar neighbors in the training set, i.e. a dense input space. Hence, the nearest neighbors' corresponding output values should more or less represent the conditional distribution of the possible outputs given the input vector in question (periodic time series, as the ones utilized in the present work, help to assure in some degree the correctness of such an assumption). The present contribution uses such property in order to change the learning pairs depending on the $q$-quantile regression ($q \in [0.01;0.99]$) which is going to be created. The value $q$ represents the nominal coverage of the desired quantile regression, meaning that if $q = 0.8$ and the regression is perfect then $80\%$ of all $y$-values will always be less than the values of the obtained model.

The procedure starts by calculating the distances between the $i^{th}$ learning pair's input and all the others, 
\begin{equation}
d_{ij} = d(\mathbf{x}_{i},\mathbf{x}_{j}) \, \forall i,j \in [1;N] \text{ ,}
\end{equation}
with $d(\cdot~,~\cdot)$ as the distance function utilized for their calculation (which can change depending on the model's application). The distance function utilized in the present work is described in one of the following sections.

The obtained distances allow the determination of the $i^{th}$ input's k-nearest-neighbors (k-NN), whose corresponding output values are sorted in ascending order. Thereafter, the values in the sorted k-NNs output set are set equal to the 
\begin{equation}
\left(\dfrac{0.5}{k_{NN}}, \dfrac{1.5}{k_{NN}}, ..., \dfrac{k_{NN}-0.5}{k_{NN}}\right)
\end{equation}
quantiles, with $k_{NN}$ being the number of used nearest-neighbors. If the desired $q$-quantile is equal to one of those, then its value is taken from the sorted k-NNs output set, if not its value is interpolated from the nearest-neighbors' output set - the described procedure is given in the works of Hyndman~\cite{Hyndman96} (defined as Definition 4) and Langford~\cite{Langford06} (referred to as Method 10). Afterwards, the $i^{th}$ output value is substituted by the obtained quantile. The process is repeated until a new set of desired outputs, $\mathbf{y}_{q}$, is created: 
\begin{equation}\label{eq:new_desired_output_vector}
\mathbf{y}_{q} =  \begin{pmatrix}
				y_{q,1}\\
				\vdots\\
				y_{q,N}
			\end{pmatrix} \text{ .}
\end{equation} 
Data mining techniques can then be trained using the changed training set and conventional regression approaches, thus delivering a model that approximates in average the conditional quantile of the desired outputs given the used input values and defined by the amount of the utilized nearest-neighbors. The above described methodology has the advantage of obtaining a functional relation for the wanted conditional quantiles using only the nearest-neighbors during the training procedure and hence, eliminating the necessity of saving the training set and of conducting the nearest-neighbors calculation during the models' usage. 
%\begin{equation}\label{eq:new_desired_output_vector}
%\mathbf{y}_{q} = (y_{q,1},	\cdots, y_{q,N})^{T}
%\end{equation} 

\begin{figure}[!tb]
	\centering
	\includegraphics[width=0.6\textwidth]{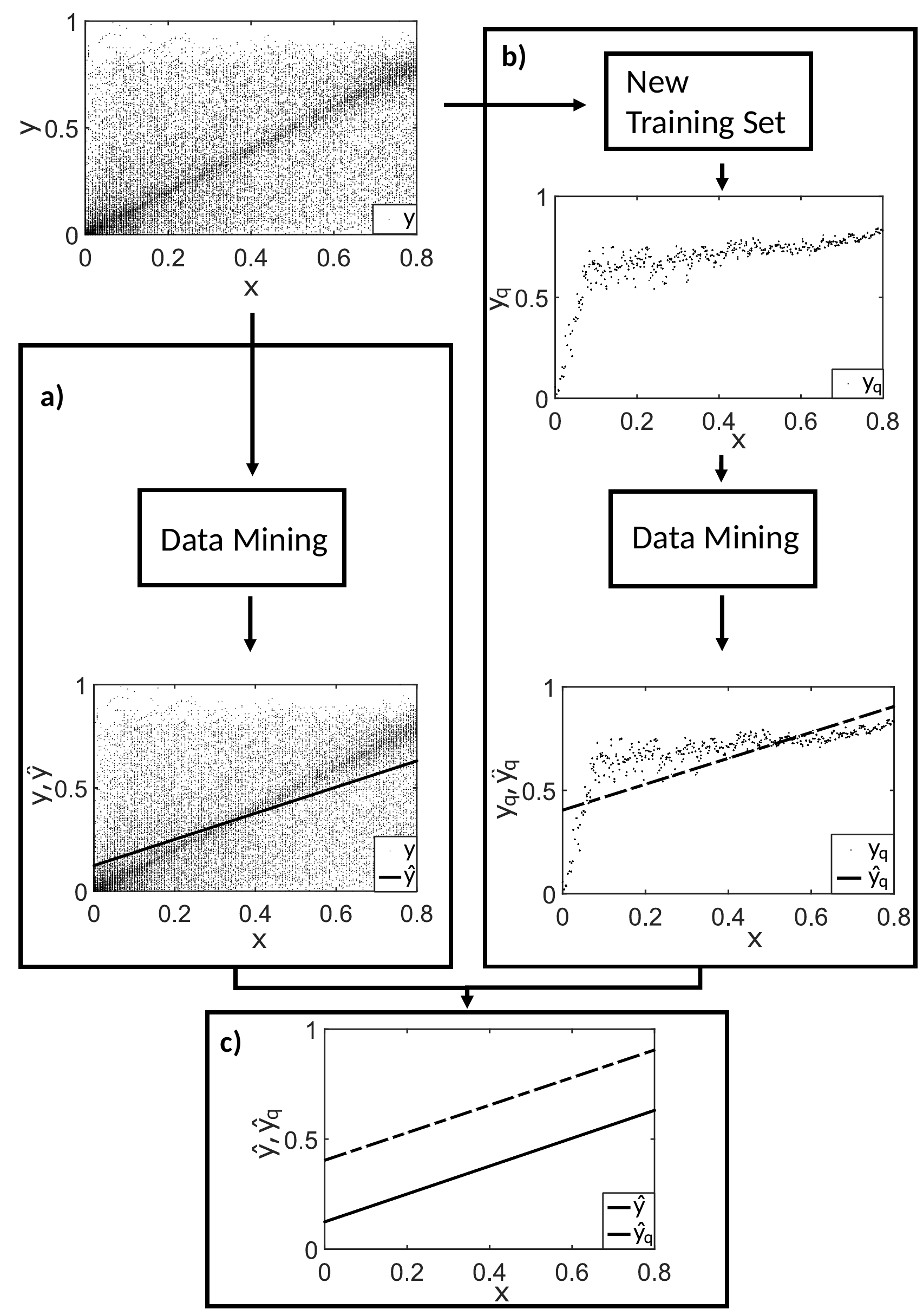}
	\caption{Nearest-neighbor based quantile regression principles using linear regression models\\
	\footnotesize	
	a) Regression model obtained using the unchanged training set\\
	b) Regression model obtained using the changed training set for $q=0.9$ \\
	c) Comparison of the two different regression models}
	\label{fig:k-NN_Quanreg_Principles}
\end{figure}

For the sake of illustration, Figure~\ref{fig:k-NN_Quanreg_Principles} shows the principles behind the present contribution's method for the case of a one dimensional input vector and a linear regression model.

\subsection{Quantile Regression Intervals}

By combining pairs of quantile regressions, intervals with a desired nominal coverage can be created. For example, an interval with a nominal coverage of $0.8$, i.e. an interval in which $80\%$ of all $y$-values will lay inside (if the interval model is perfect), can be obtained by using the $0.9$ and the $0.1$-quantile regressions; with the former as the upper interval bound ($\hat{y}_{q_{u}}$) and the latter as the lower interval bound ($\hat{y}_{q_{l}}$). The combination of both quantile regressions is given as:
\begin{equation}
\mathbf{\hat{y}}_{(q_{u}-q_{l})} = \begin{bmatrix}
         \hat{y}_{q_{u}} \\
         \hat{y}_{q_{l}}
         \end{bmatrix} \text{ ,}
\end{equation}
with $(q_{u}-q_{l})$ representing the difference between the nominal coverage of the upper and lower bounds and thus, the nominal coverage of the created interval.

\begin{figure}[!tb]
\centering
\begin{subfigure}[b]{0.49\textwidth}
        		\centering
	\includegraphics[width=.8\textwidth]{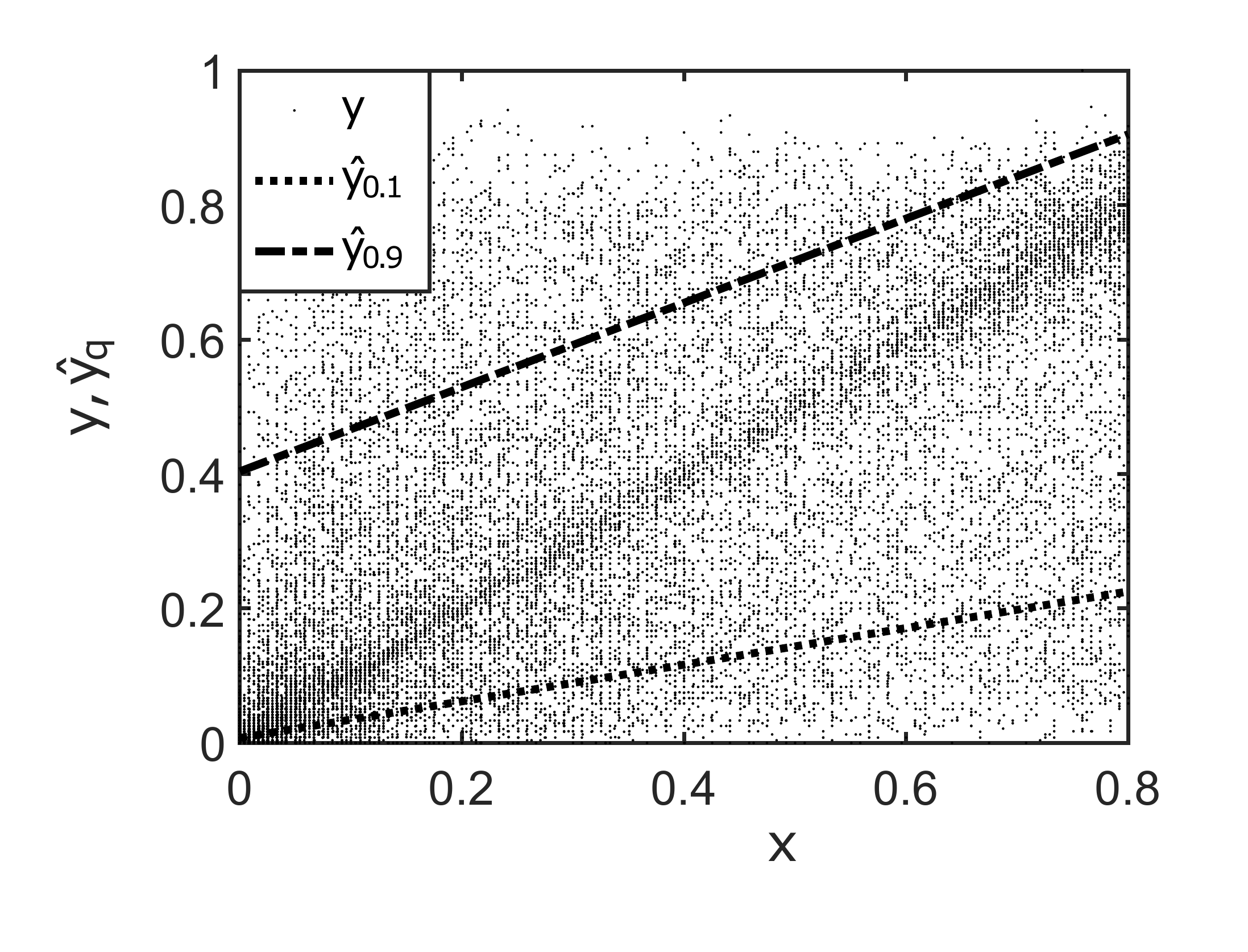}
	\caption{Quantile regression in feature space}
	\label{fig: example_featureSpace}
\end{subfigure}
\begin{subfigure}[b]{0.49\textwidth}
        		\centering
	\includegraphics[width=.8\textwidth]{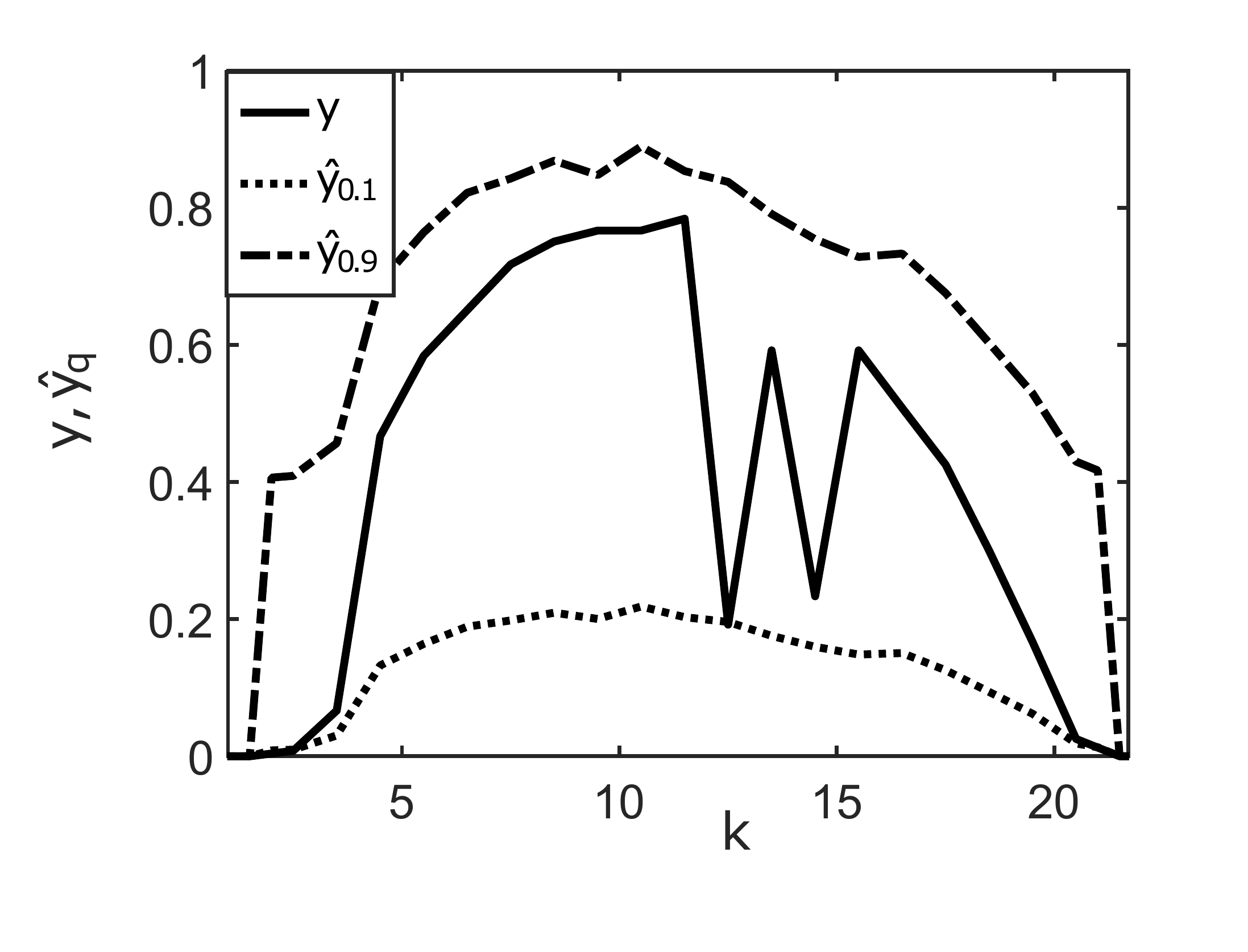}
	\caption{Quantile regression as time series}
	\label{fig: example_timeSeries}
\end{subfigure}
\caption{Example of interval forecast using pairs of linear quantile regressions}
\label{fig: example_featureSpace_timeSeries}
\end{figure}

In the present contribution, the regressions' desired output values are future values of a time series, so the created intervals can be utilized as interval forecasting models.
An illustrative example of linear quantile regressions forming the upper and lower bounds, when represented as regression models (for a one-dimensional input) and as time series, is shown in Figure~\ref{fig: example_featureSpace_timeSeries}. 
%\begin{figure}[!htb]
%	\centering
%	\includegraphics[width=\textwidth]{Images/k-NN_Quanreg_FeatureSpace-TS.eps}
%	\caption{Example of Interval Forecast using Pairs of Quantile Regressions \\
%	a) Quantile Regressions in Feature Space \\
%	b) Quantile Regressions as Time Series}
%	\label{fig:k-NN_QuanReg_FeatureSpace-TS}
%\end{figure}

\subsection{Evaluation}

Traditional evaluation values like mean absolute error or root mean square error are not adequate to determine the accuracy of quantile regressions and the respectively created intervals. Therefore, other kinds of evaluation values have been proposed in literature.

The first value used in the present work is the reliability~\cite{Hong16a,Juban07,Pinson07,Zhang14}. It describes the actual coverage of a quantile regression or an interval, i.e. the percentage of values which actually lay under the former and inside the latter. Due to the fact, that a common way of depicting the accuracy of quantile regressions or intervals is by plotting their reliability deviation (i.e. the difference between their actual coverage and their nominal coverage) against their nominal coverage, the reliability deviation $\Delta Rl$ is chosen as the present contribution's evaluation value. Such value is given for a single quantile regression and a testset with N samples as
\begin{equation}\label{eq: reliability_dev_quanReg}
\Delta Rl_{q}  =  \dfrac{1}{N}\underset{n=1,..,N}{\operatorname{card}}(y_{n} < \hat{y}_{q,n}) - q  \text{ ,} 
\end{equation}
while for an interval formed by quantile regressions as
\begin{equation}
\Delta Rl_{(q_{u}-q_{l})} = \dfrac{1}{N}\underset{n=1,..,N}{\operatorname{card}}(\hat{y}_{q_{l},n} \leq y_{i} < \hat{y}_{q_{u},n}) - (q_{u}-q_{l}) \text{ ;} \\	
\end{equation}
with $\operatorname{card}(\cdot)$ being the cardinality operator. The more the reliability deviation approximates zero, the better the evaluated model is. A negative value describes an underestimating quantile regression or a narrow interval, while a positive value points at the contrary. Furthermore, the reliability deviation does not quantify the extent in which the measurements violate the quantile regression or the interval bounds, but instead treats every violation equally. A problem that arises by using the reliability deviation is, that if a quantile regression or an interval underestimate in some cases, but overestimate in others by the same amount, then their reliability deviation will result in a perfect regression or interval. However, such result is incorrect.

Due to the above mentioned problems other evaluation values presented by Gneiting et. al~\cite{Gneiting14} are also used in the present work. First, in order to evaluate the quality of quantile regressions the pinball-loss is used. The pinball-loss for a single quantile regression and a testset with N samples is given as 
\begin{equation}\label{eq: pinball_loss_quantreg}
L_{q} = \underset{n=1,..,N}{\operatorname{mean}}\{(y_{n} - \hat{y}_{q,n})(q-I(y_{n}<\hat{y}_{q,n}))\} \text{ ,}
\end{equation}
with $I(\cdot)$ being the indicator function\footnote{Function which equals 1 if its condition is fulfilled and 0 otherwise.}. As it can be seen in Equation~(\ref{eq: pinball_loss_quantreg}) the pinball-loss considers the magnitude of each deviation and weights it depending on the quantile regression being evaluated. The lower its value is, the better the evaluated regression is. A similar value, referred in the present work as interval's pinball-loss, is used for the evaluation of the created intervals. Such value can only be utilized if the condition $q_{u} = 1-q_{l}$ holds and it is given as 
\begin{equation}\label{eq: interval_pinball_loss}
\begin{aligned}
 L_{(q_{u}-q_{l})} = & \underset{n=1,..,N}{\operatorname{mean}}\{(\hat{y}_{q_{u},n}-\hat{y}_{q_{l},n}) +  \\
                     & \dfrac{2}{1-(q_{u}-q_{l})}(y_{n} - \hat{y}_{q_{u},n})I(y_{n}>\hat{y}_{q_{u},n}) + \\
 					 &  \dfrac{2}{1-(q_{u}-q_{l})} (\hat{y}_{q_{l},n} - y_{n})I(y_{n}<\hat{y}_{q_{l},n}) ) \} \text{ .}
\end{aligned}
\end{equation}
The interval's pinball-loss has as its first term the distance between its bounds, thus considering a narrow interval better than a broader one and allowing the identification of unwanted trivial intervals. A trivial interval is, for example, an interval with a nominal coverage of $0.99$ for $y$-values ranging between $0$ and $1$ formed by the two constant values $0$ and $0.99$. The second and third term quantify the amount in which the values outside of the interval deviate from it; those deviations are weighted depending on the interval's nominal coverage. The lower the pinball-loss value is, the better the evaluated interval is.

\section{Photovoltaic Power Forecasting}

The goal of the present contribution's probabilistic photovoltaic power forecasting is to obtain accurate quantile regressions with $q \in [0.01;0.99]$ for a forecast horizon of $24h$ utilizing information of the past generated power, as well as the previously described nearest-neighbor based approach. Those regressions are obtained with various amounts of nearest-neighbors ($50,70,100,$ and $120$) for the training set transformations. After their creation they are combined into probabilistic forecasting models. The present section describes the data and techniques utilized for creating the presented PV power forecasting models.

\subsection{Dataset}

The dataset utilized in the present contribution is a freely available dataset provided by the Australian energy provider Ausgrid\footnote{http://www.ausgrid.com.au}. After the dataset's preprocessing (which is conducted just as in~\cite{GonzalezOrdiano16}), time series containing measurements of the normalized average generated photovoltaic power (normalized to values between $0$ and $1$) every 15 minutes for the time frame of July $1^{st}$, 2010  to June $30^{th}$, 2013 ($K=52608$) of 54 rooftop PV systems are obtained. The 54 rooftop PV systems comprise the clean dataset defined by Ratnam et al.~\cite{Ratnam15a}. The dataset does not contain historical nor forecast weather information, hence all forecasting models described in the following sections are created using purely the information of the past generated power. Furthermore, the utilized time series are separated in halves, with the first half used as training set and the second as test set.

\subsection{Input Time Series}

As already mentioned, the models shown in the present contribution only utilize information of the past generated power ($P$). From $P$, two other time series, which are also utilized as input, are created. The first one is given by the equation
\begin{equation}
P_{max}[k] = \operatorname{max}\{P[k],P[k-H_{p}],...,P[k-m \cdot H_{p}]\} \text{ ,}
\end{equation}   
with $H_{p}$ being the number of time steps representing $24$ hours (i.e. the day's periodicity) and $m = 7$. This new created time series contains the measured maximal values of the last eight days. Additionally, the second created time series, given by the equation
\begin{equation}
P_{mean}[k] = \operatorname{mean}\{P[k],P[k-H_{p}],...,P[k-m \cdot H_{p}]\} \text{ ,}
\end{equation} 
contains the average values of the last eight days. The creation of the previously described time series has the goal of eliminating the time series' random effects, while retaining systematic repeating ones (e.g. shadowing effects caused by neighboring buildings), with $P_{mean}$ also retaining information of the previous week's weather variability. %The utilization of both $P_{max}$ and $P_{mean}$ as additional model inputs has the goal of reducing the influence that random effects have on the interval forecasting models, due to the fact that its happening at the forecast horizon is extremely unlikely. The complete opposite can be said of systematic repeating effects. Hence it is important to retain them. The exclusion of a time series containing the minimal values of the last eight days comes from the fact that such a time series appeared to possess a great amount of random effects, e.g. cloud coverages, which could reduce the quality of the interval forecasts. 
The usage of maximal and mean time series to reduce the influence of past random effects is reserved for periodic time series, due to their repeating nature. %Furthermore, both $m$ and $H_{1}$ can be changed depending on the periodicity of the utilized time series (e.g. weekly periodicity of load time series).

With this time series, Equation~(\ref{eq. quantile_forecast_functional_relation}) is extended to
\begin{equation}\label{eq:equation forecast horizon 24h}
\begin{aligned}
\hat{P}_{q}[k+H] = & f_{q}(P[k],..,P[k-H_{1}],P_{max}[k],..,P_{max}[k-H_{1}],\\
                & P_{mean}[k],..,P_{mean}[k-H_{1}];\boldsymbol{\theta}_{q}) \text{ .}
\end{aligned}
\end{equation}
In the present work, the values $H$, $H_{p}$, and $H_{1}$ are defined as $96$ time steps (number of timesteps representing 24h due to the utilized time series' resolution).
%Further values of other time series (e.g. forecast solar irradiation) can be added as inputs, but it has to considered that with an increasing number of input values the probability of having a dense dataset, necessary for the present contributions method, decreases. 
%As a next step, a feature selection is applied, hence, reducing the dimensionality of the input space and the risk of sparsity.

\subsection{Elimination of Night Values}

In order to increase the accuracy of the probabilistic PV power forecasting models, night values are eliminated from the used training set. For the elimination, the following assumption is made: If the generated PV powers 24 and 48 hours prior to the forecast horizon ($H=96$) are less or equal to a threshold, then both values as well as the value at the forecast horizon are most likely night values; thus both the value at the forecast horizon (desired output) and its corresponding input vector are eliminated from the training set. This assumption allows the creation of the forecasting models on the basis of non-trivial values. In the present contribution the threshold was set equal to $0.0001$. During the quantile regressions' usage the same assumption is utilized. Therefore, only future values which are considered to be day values are forecast by the created models, all other values are set automatically equal to zero (night values). Likewise, the quantile regressions and the interval forecasting models are only evaluated on test set values considered to be day measurements. 
%Furthermore, during the quantile regressions' usage, the following equation is applied
%\begin{equation}
%P_{q}[k+H]=\begin{cases}
%			     		0                            & \text{, if } P[k] \leq n_{th} \wedge P[k-H_{1}] \leq n_{th} \\
%						f_{q}(P,P_{max},P_{mean}) & \text{, else} \\								
%						\end{cases} \text{ ,} ($n_{th}$)
%\end{equation}
%thus applying the models only to forecast values which are considered day values by the utilized assumption. Additionally, the quantile regressions and the interval forecasting models %are only evaluated on values considered to be day measurements. In the present contribution $n_{th}$ was set equal to $0.0001$.

\subsection{Data Mining Techniques}

The present contribution utilizes three different polynomial models without bilinear terms, two artificial neural networks, and a support vector regression to determine quantile regressions for $q \in [0.01;0.99]$ with which the PV power interval forecasts are to be obtained. The polynomial models are referred to as Poly1, Poly2, and Poly3 and describe polynomials with a maximal degree of one, two, and three. The artificial neural networks (ANN) are multilayer perceptrons with a single hidden layer, but a different number of neurons in it; the first referred to as ANN6 has six neurons, while the second, ANN10, has ten neurons. Finally, the support vector regression (SVR) utilizes a Gaussian kernel of degree one, a trade off value $C$ equal to $1$ and an $\epsilon$ equal to $0.01$. All utilized models are created using the MATLAB toolbox Gait-CAD~\cite{Mikut08Biosig}.

Before the calculation of the PV power quantile regressions is undertaken, the distance function used for the determination of the nearest-neighbors has to be defined. In the present work, the weighted Euclidean distance is utilized. The necessary weights are defined as the inverse of the features' variance in the used training set. Both the distance function and its weights are described as
\begin{equation}\label{eq: weighted_euclidean}
\begin{aligned}
& d_{ij} = \left(\sum_{s=1}^{S}w_{s}(x_{is}-x_{js})^{2}\right)^{1/2} \, \forall i,j \in [1;N] \text{ ,} \\
& w_{s} =  (\operatorname{var}(\mathbf{X}(:,s))^{-1} \text{ for } s \in[1,S] \text{ .}
\end{aligned}
\end{equation}
For features which are constant, a regularization has to be conducted during its weight calculation in order to avoid dividing by zero.

Furthermore, from all input values shown in Equation~(\ref{eq:equation forecast horizon 24h}), four of them are selected individually for each model type and each household to be utilized as features during the creation and application of the models, hence $S=4$. The selection process assures that the only difference between the quantile regressions from a specific technique are not the utilized features, but the parameters $\boldsymbol{\theta}_{q}$ obtained from their different training sets. Additionally, the selection of four features prior to the nearest-neighbors calculation reduces the possibility of sparsity in the used feature space, which in high dimensions would be inevitable - even when utilizing the past values of the periodic PV power time series as input. The selection procedure is a forward feature selection. Due to the long computational time required for the SVR models' creation, the forward feature selection is not applied for this type of model; instead, the features selected for ANN6 are used. According to the notation in Equations~(\ref{eq:input_matrix}) and (\ref{eq:new_desired_output_vector}) the input vectors and their corresponding desired outputs, necessary for the application of the present work's method, are given as $\mathbf{x}_{n}^{T} = (x_{n1},x_{n2},x_{n3},x_{n4})$ and $y_{q,n} = P_{q}[k+H]$.

The actual obtainment of the PV power polynomial quantile regressions starts by determining the polynomial describing the smallest quantile ($q = 0.01$) by minimizing the sum of squared errors and using the constraint that its values should be greater than zero. Afterwards, the remaining quantile regressions are created under the constraint that their values should be greater than the ones obtained by the previously obtained quantile regression model. This constraint avoids the problem of quantile crossing~\cite{Fahrmeir09}, in which a quantile regression delivers values which are smaller than the ones provided by a model representing a lower quantile. Additionally the vector $\mathbf{1}_{N}$, which only contains ones, is added to the input matrix, hence allowing offsets in the desired polynomials, thus making them affine functions. Nonetheless, the distances in Equation~(\ref{eq: weighted_euclidean}) are still calculated using the original input matrix $\mathbf{X}$. The determination of the polynomials is given by the following equations:
\begin{equation}
\begin{aligned}
& \underset{\boldsymbol{\theta}_{q}}{\operatorname{minimize}} & ||\mathbf{y}_{q} & - [\mathbf{1}_{N} ~\mathbf{X}]\boldsymbol{\theta}_{q}||^{2}\\
& \operatorname{subject~to} & & \begin{cases} 
								  [\mathbf{1}_{N}~\mathbf{X}]\boldsymbol{\theta}_{q}  \geq 0 & \text{ if $q$ = 0.01}\\
                                  [\mathbf{1}_{N}~\mathbf{X}]\boldsymbol{\theta}_{q} \geq [\mathbf{1}_{N}~\mathbf{X}]\boldsymbol{\theta}_{(q-0.01)} & \text{ else}  \end{cases}\\
& \operatorname{with} & \boldsymbol{\theta}_{q}^{T} & = ( \theta_{q,0},\theta_{q,1},\theta_{q,2},\theta_{q,3},\theta_{q,4}) \text{ .}
\end{aligned} 
\end{equation}
The greater than zero constraint introduced for the smaller quantile stems from the a-priori known fact that generated PV power should only be positive. The possibility of implementing all of the constraints with ease is the main reason why polynomial models are utilized in the present contribution. Of course, their implementation comes at the cost of making the assumption that the output conditional quantiles must change linearly in the given feature space.

The creation of the ANN6 and ANN10 models is conducted utilizing the Levenberg–Marquardt algorithm with a maximum of $20$ training epochs, while the SVR models are obtained with the Gait-CAD implementation of the libsvm C++ library. The ANNs and SVR do not require the linearity assumption of the polynomials. However, they have the drawback of not using any constraints during the creation of each quantile regression. Therefore, in order to avoid quantile crossings, the following equation, with $\hat{y}_{q,n}=\hat{P}_{q}[k+H]$, has to be used during their evaluation:
\begin{equation}
\hat{\tilde{y}}_{q,n} = \begin{cases}
				\operatorname{max}(\hat{y}_{q,n}, 0) & \text{ if $q = 0.01$} \\
				\operatorname{max}(\hat{y}_{q,n}, \hat{y}_{q-0.01,n}) & \text{ else}			
				\end{cases} \text{ .}
\end{equation}
The value $\hat{\tilde{y}}_{q,n} = \hat{\tilde{P}}_{q}[k+H]$ represents the corrected value after the application of the above equation.

\section{Results}

First of all, the quality of the present work's assumption to identify night values has to be addressed. It shows an accuracy of $99\%$ when tested on the complete utilized dataset. 

In order to evaluate the different quantile regressions, averages of the reliability deviation's absolute values and the pinball-loss on the utilized test set are presented in Tables~\ref{tab:reliability_deviation} and ~\ref{tab:pinball_loss}. The given values are averages across the $54$ used households and $99$ quantile regressions created with each data mining technique and different amounts of nearest-neighbors. Furthermore, those values are obtained by evaluating the models on the test set values considered to be day measurements. 
\begin{table}[tb]
  \caption{Average quantile regressions' reliability deviation on test set for $H=96$}  
  \centering
  \begin{tabular}{c|ccccccc}
	\toprule
	& Poly1 & Poly2 & Poly3 & ANN6 & ANN10 & SVR \\
    $k_{NN}$ & $\%$ & $\%$ & $\%$ & $\%$ & $\%$ & $\%$\\    
    \midrule
     50  & 5.93 & 5.66 & 5.89 & 4.44 & 4.24 & 1.60\\
     70  & 5.95 & 5.68 & 5.95 & 4.28 & 3.87 & 1.59\\
     100 & 5.95 & 5.67 & 5.93 & 4.07 & 3.61 & 1.63\\
     120 & 5.94 & 5.66 & 5.91 & 3.81 & 3.47 & 1.68\\
    \bottomrule
    \end{tabular}
  \label{tab:reliability_deviation}
\end{table}
\begin{table}[tb]
  \caption{Average quantile regressions' pinball-loss on test set for $H=96$}  
  \centering
  \begin{tabular}{c|ccccccc}
	\toprule
	& Poly1 & Poly2 & Poly3 & ANN6 & ANN10 & SVR \\
    $k_{NN}$ & $\%$ & $\%$ & $\%$ & $\%$ & $\%$ & $\%$\\    
    \midrule
     50  & 4.21 & 4.21 & 4.22 & 4.14 & 4.14 & 4.11\\
     70  & 4.21 & 4.21 & 4.22 & 4.14 & 4.14 & 4.11\\
     100 & 4.21 & 4.21 & 4.22 & 4.14 & 4.14 & 4.12\\
     120 & 4.21 & 4.21 & 4.22 & 4.13 & 4.14 & 4.12\\
    \bottomrule
    \end{tabular}
  \label{tab:pinball_loss}
\end{table}

\begin{figure}[!tb]
\centering
\begin{subfigure}[b]{0.49\textwidth}
        		\centering
	\includegraphics[width=.8\textwidth]{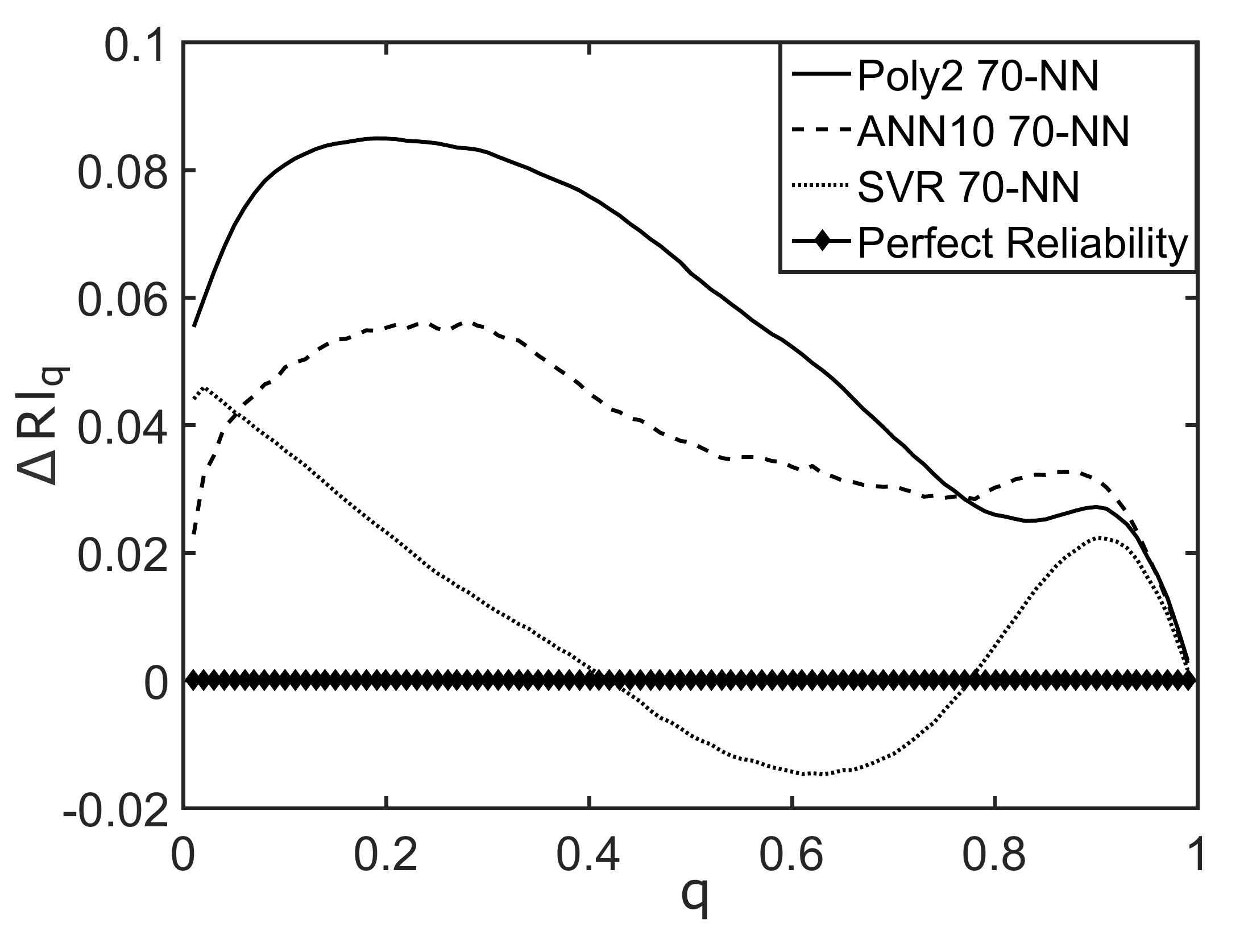}
	\caption{Poly2, ANN10, and SVR with 70-NNs}
	\label{fig: reliability_deviation_70NN}
\end{subfigure}
\begin{subfigure}[b]{0.49\textwidth}
        		\centering
	\includegraphics[width=.8\textwidth]{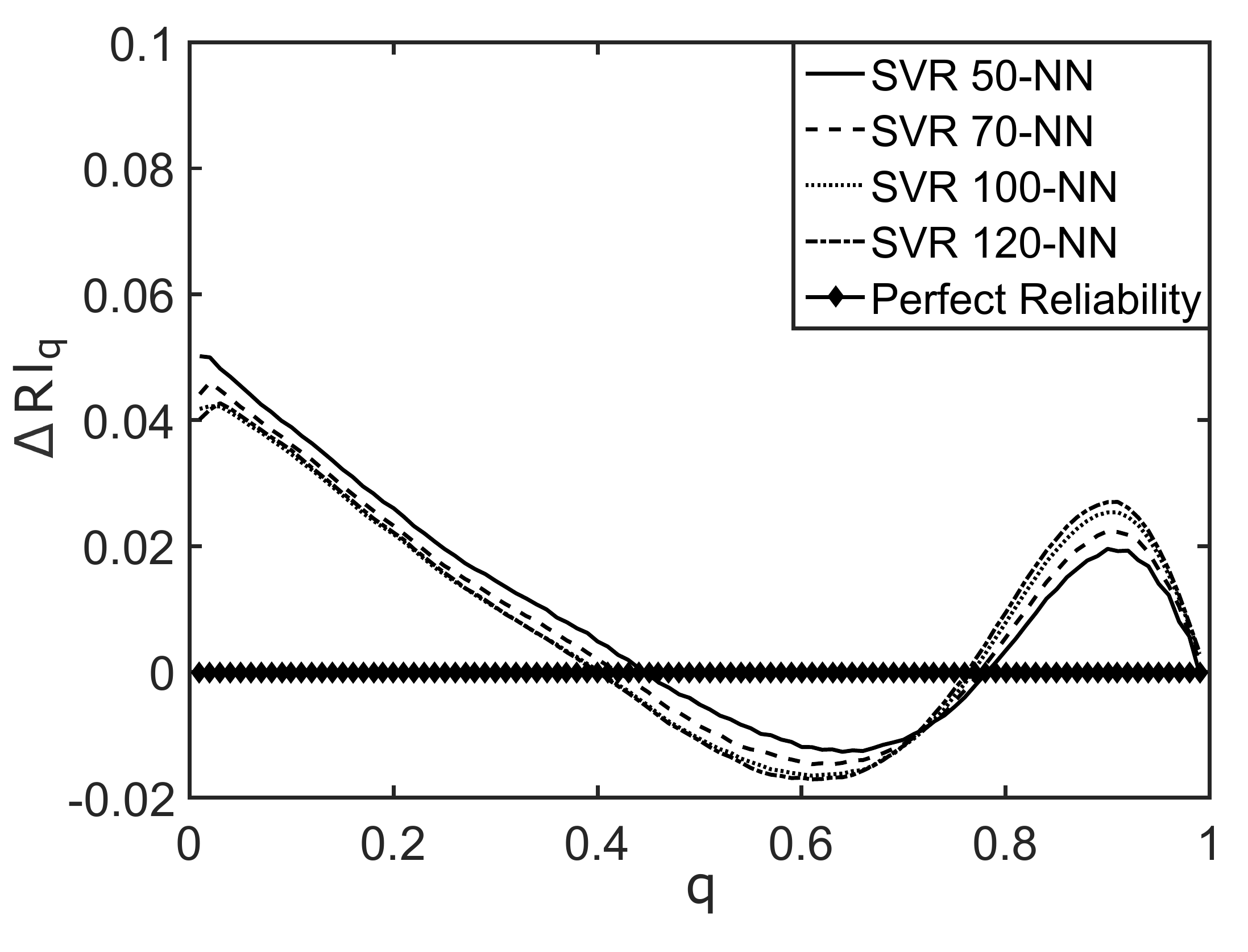}
	\caption{SVR models with varying NNs' amount}
	\label{fig: reliability_deviation_SVR}
\end{subfigure}
\caption{Test set reliability deviations of quantile regressions}
\end{figure}

The contents in Table~\ref{tab:reliability_deviation} show a clear improvement in the reliability deviation when utilizing the non-linear techniques, with SVR - especially SVR with $k_{NN}=70$ - showing the lowest average reliability deviations. The average pinball-loss values in Table~\ref{tab:pinball_loss} also show the improvements of utilizing non-linear approaches. Nonetheless, the differences in the pinball-loss averages are not as remarkable as the ones shown in the reliability deviation results. The SVR with $70$-NNs creates in average the best quantile regressions, for such reason, Figure~\ref{fig: reliability_deviation_70NN} compares the reliability deviation for each nominal coverage to the ones obtained by the best polynomial (Poly2) and the best artificial neural network (ANN10) for $k_{NN}=70$. Likewise Figure~\ref{fig: reliability_deviation_SVR} plots the reliability deviation against the nominal coverage for the SVR models obtained using different amounts of NNs. 

Figure~\ref{fig: reliability_deviation_70NN} shows that SVR with $k_{NN}=70$ obtains the models whose absolute reliability deviation is the lowest; with the only exception being the lowest nominal coverages in which ANN10 delivers regressions with a lower deviation. Likewise, it is possible to discern that both Poly2 and ANN10 overestimate in average, while SVR fluctuates between overestimating and underestimating depending on the considered nominal coverage. Interestingly, the non-linear ANN10 shows an advantage over Poly2 until a nominal coverage of approximately $0.8$, afterwards, Poly2 becomes better. Figure~\ref{fig: reliability_deviation_SVR} depicts how the increasing number of nearest-neighbors affects the accuracy. The figure shows that the lower the amount of nearest-neighbors is, the greater the reliability deviation for low quantiles (for $q$ lower than $0.4$) is. However, for higher quantiles (for $q$ higher than $0.8$) the opposite is true. The effect could be explained by the fact that using a higher amount of nearest-neighbors allow the lower quantiles to reflect effects like complete changes in weather (e.g. like rainy days after a couple of sunny days), but at the same time introduce output values corresponding to not so similar input vectors that skew the higher quantile regressions. 

By combining pairs of the obtained quantile regressions (starting with the regressions for $q_{u} = 0.51$ and $q_{l} = 0.49$ and ending with $q_{u} = 0.99$ and $q_{l} = 0.01$) $49$ different interval forecasting models, with nominal coverage from $0.02$ to $0.98$, are created. Table~\ref{tab:reliability_deviation_interval_forecast} contains the reliability deviation average across all nominal coverage for the techniques' interval forecasting models. Likewise, Table~\ref{tab:pinball_loss_interval_forecast} contains its average interval's pinball-loss values.
\begin{table}[tb]
  \caption{Average intervals' reliability deviation on test set for $H=96$}  
  \centering
  \begin{tabular}{c|cccccc}
	\toprule
	& Poly1 & Poly2 & Poly3 & ANN6 & ANN10 & SVR \\
    $k_{NN}$ & $\%$ & $\%$ & $\%$ & $\%$ & $\%$ & $\%$\\    
    \midrule
     50  & 4.57 & 4.51 & 4.76 & 2.48 & 2.15 & 2.15\\
     70  & 4.24 & 4.21 & 4.47 & 2.31 & 1.86 & 1.84\\
     100 & 3.98 & 3.99 & 4.26 & 1.80 & 1.69 & 1.61\\
     120 & 3.86 & 3.90 & 4.19 & 1.91 & 1.58 & 1.56\\
    \bottomrule
    \end{tabular}
  \label{tab:reliability_deviation_interval_forecast}
\end{table}
\begin{table}[tb]
  \caption{Average intervals' pinball loss on test set for $H=96$}  
  \centering
  \begin{tabular}{c|cccccc}
	\toprule
	& Poly1 & Poly2 & Poly3 & ANN6 & ANN10 & SVR \\
    $k_{NN}$ & $\%$ & $\%$ & $\%$ & $\%$ & $\%$ & $\%$\\    
    \midrule
     50  & 38.56 & 38.54 & 38.66 & 37.93 & 37.95 & 37.72\\
     70  & 38.54 & 38.53 & 38.65 & 37.90 & 37.96 & 37.73\\
     100 & 38.53 & 38.52 & 38.64 & 37.91 & 37.90 & 37.77\\
     120 & 38.53 & 38.51 & 38.64 & 37.88 & 37.92 & 37.80\\
    \bottomrule
    \end{tabular}
  \label{tab:pinball_loss_interval_forecast}
\end{table}

\begin{figure}[!tb]
\centering
\begin{subfigure}[b]{0.49\textwidth}
        		\centering
	\includegraphics[width=.8\textwidth]{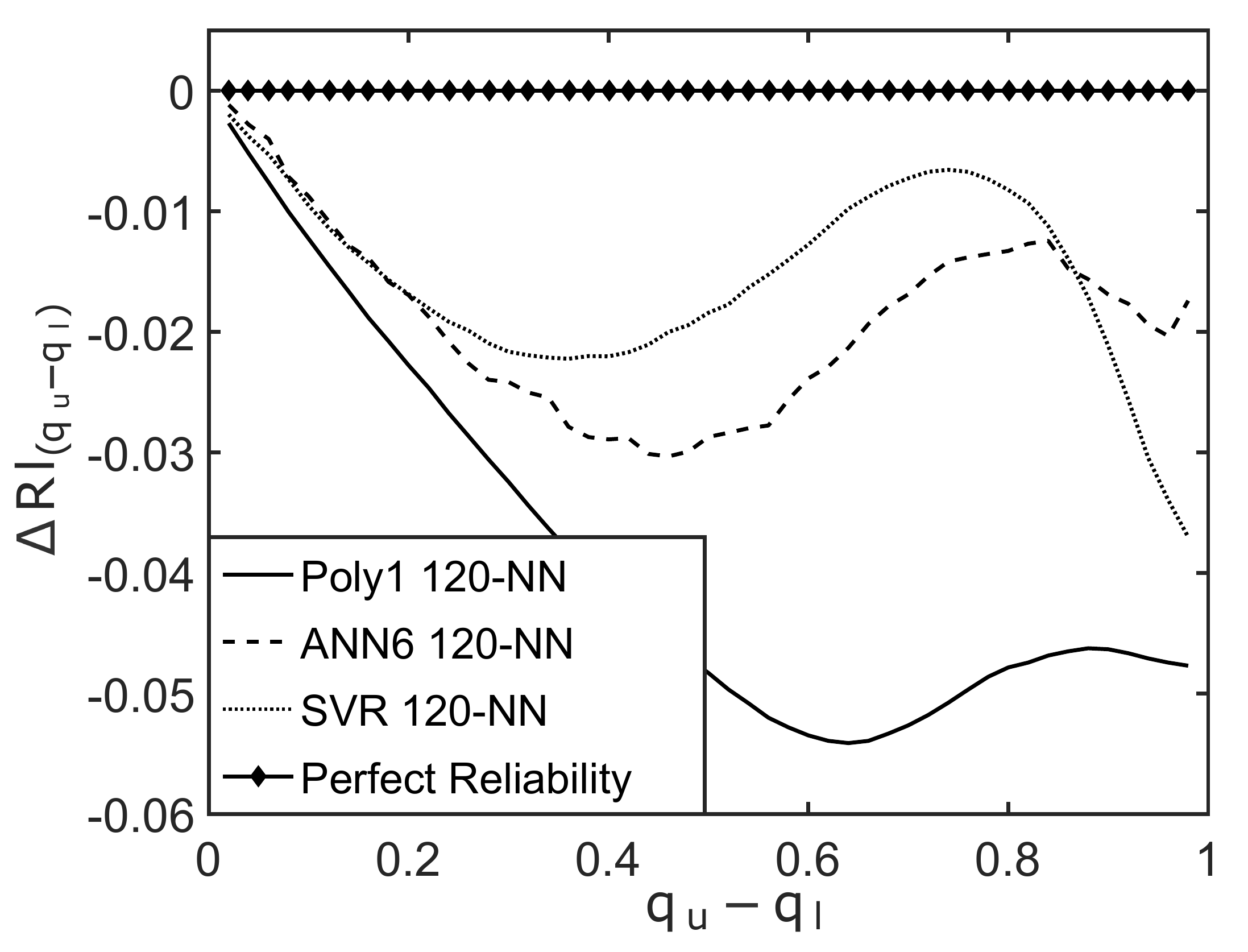}
	\caption{Reliability deviation}
	\label{fig: reliability_deviation_interval_120NN}
\end{subfigure}
\begin{subfigure}[b]{0.49\textwidth}
        		\centering
	\includegraphics[width=.8\textwidth]{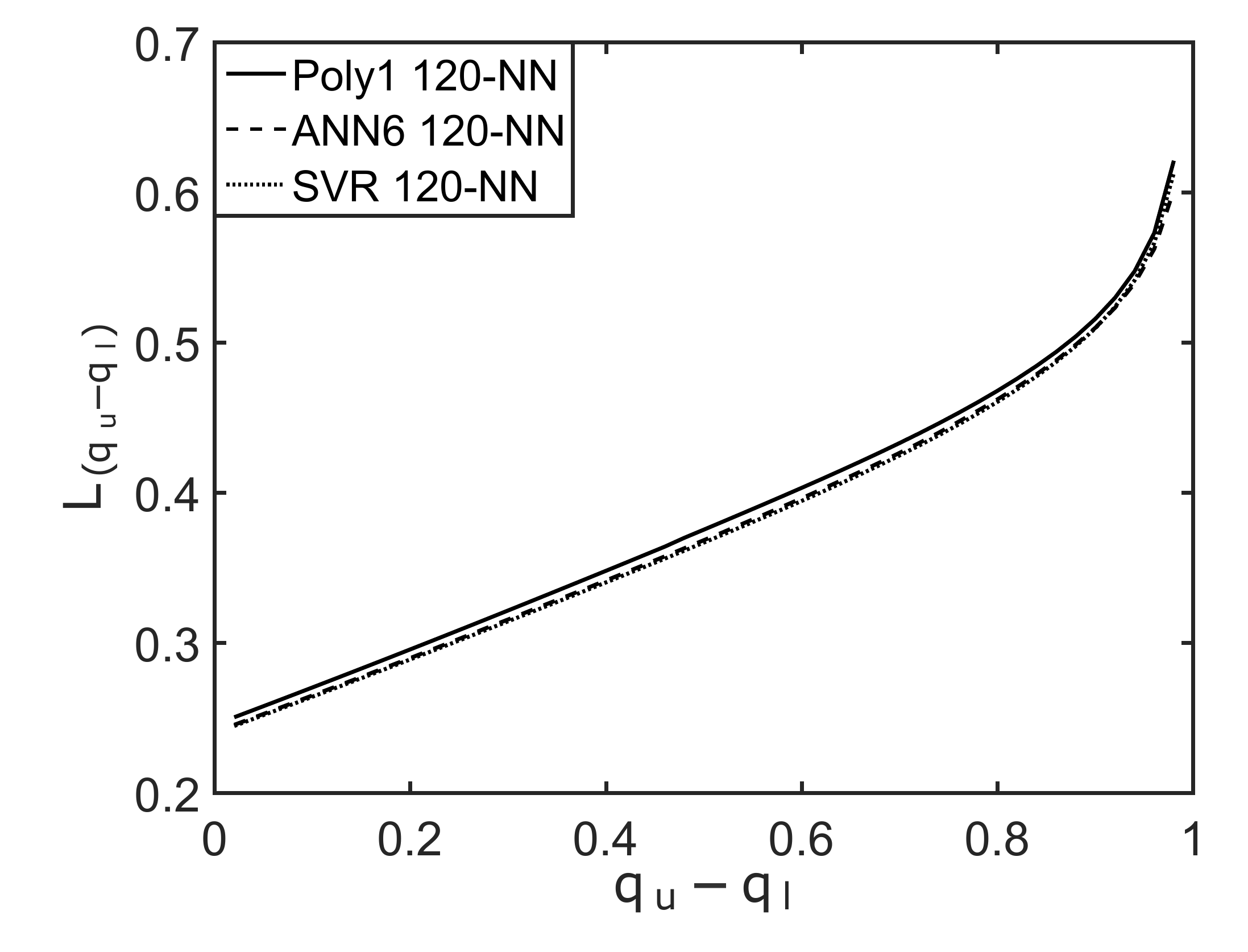}
	\caption{Interval's pinball loss}
	\label{fig: pinball_loss_Interval_120NN}
\end{subfigure}
\caption{Test set evaluation values of interval forecasts with 120-NNs}
\end{figure}

\begin{figure}[!tb]
\centering
\begin{subfigure}[b]{0.49\textwidth}
        		\centering
	\includegraphics[width=0.8\textwidth]{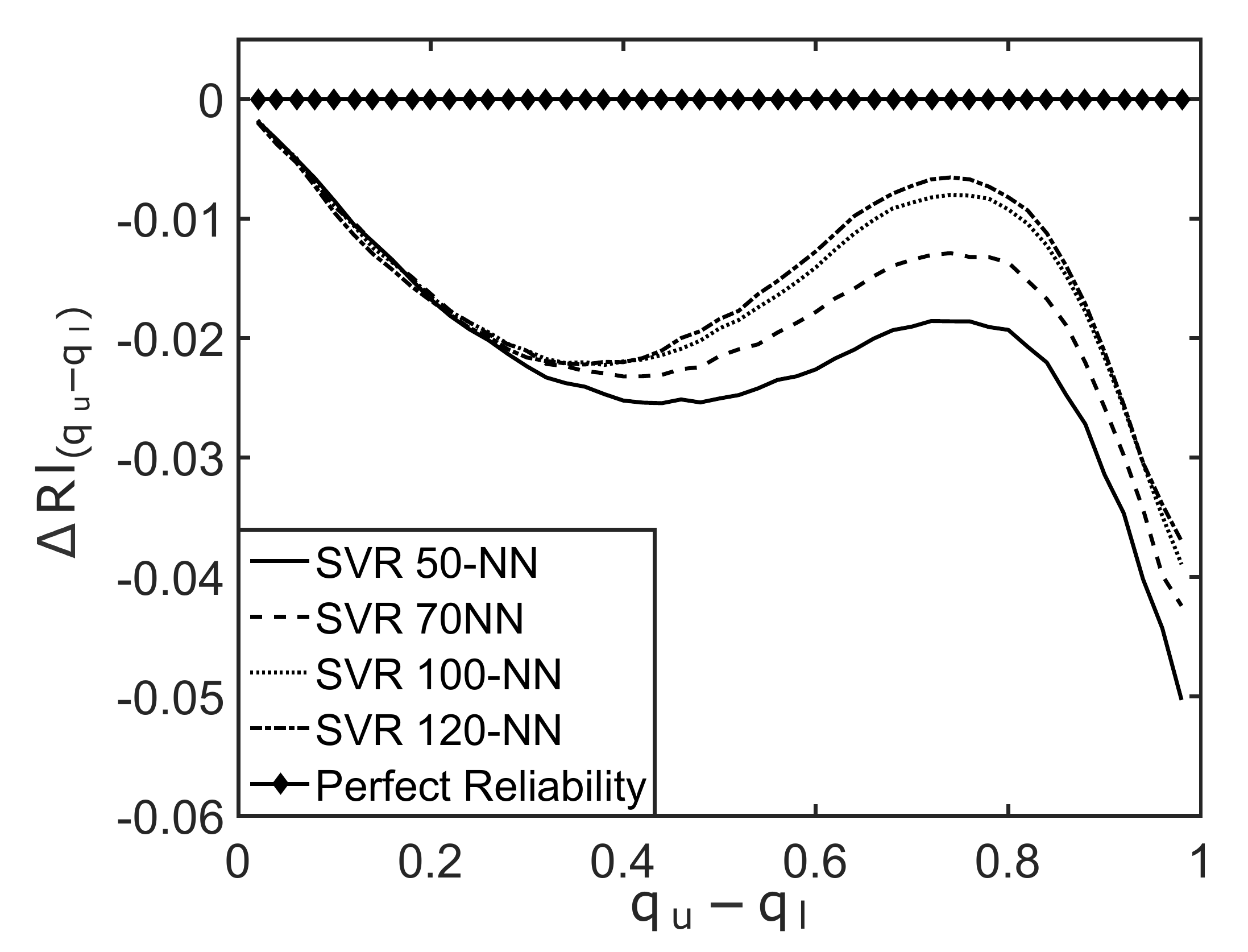}
	\caption{Reliability deviation}
	\label{fig: reliability_deviation_Interval_SVR}
\end{subfigure}
\begin{subfigure}[b]{0.49\textwidth}
        		\centering
	\includegraphics[width=0.8\textwidth]{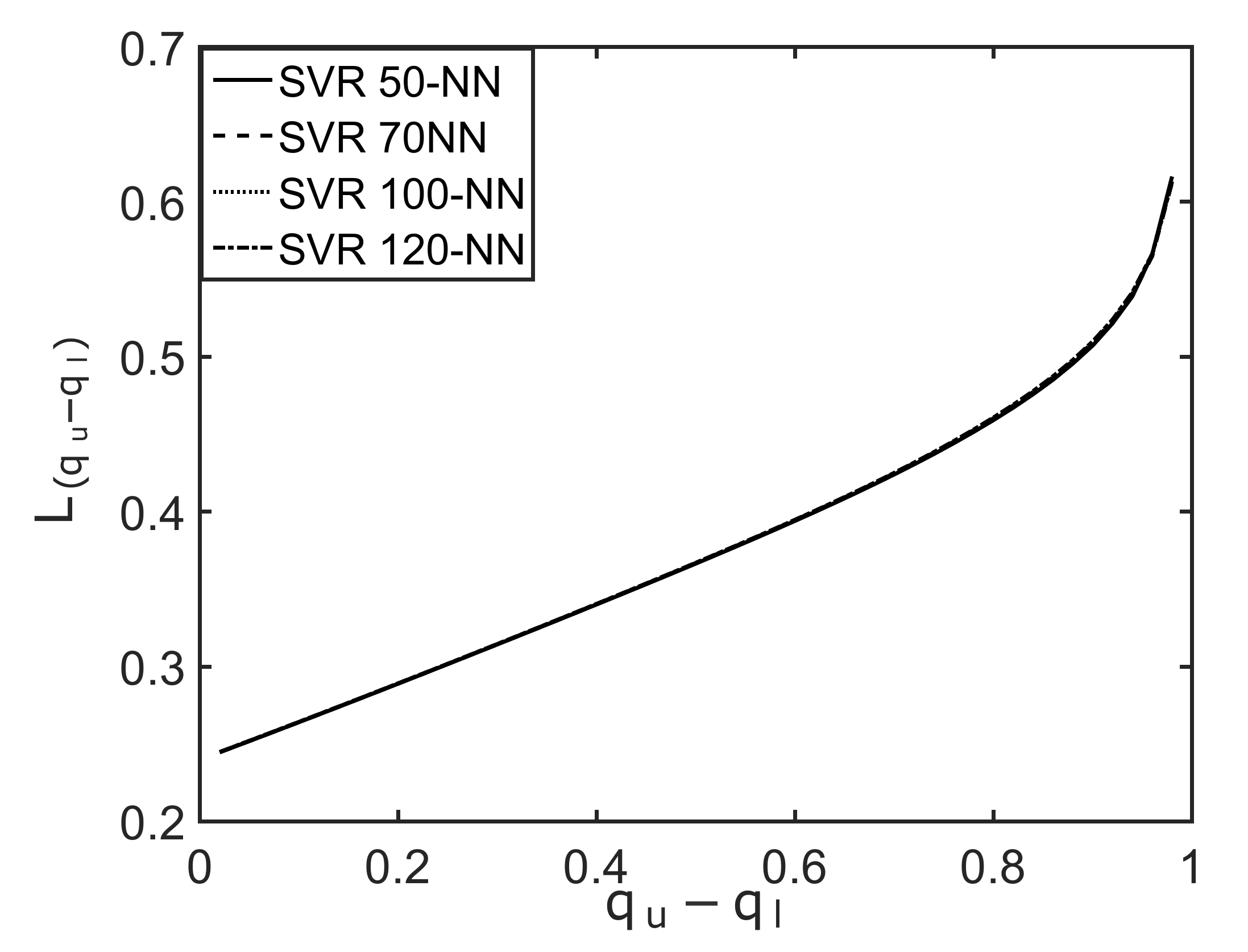}
	\caption{Interval's pinball loss}
	\label{fig: pinball_loss_Interval_SVR}
\end{subfigure}
\caption{Test set evaluation values of SVR interval forecasts}
\end{figure}

Just as before, the values contained in Tables~\ref{tab:reliability_deviation_interval_forecast} and \ref{tab:pinball_loss_interval_forecast} show an improvement through the usage of non-linear approaches. With SVR again as the technique delivering the most accurate models. The one using 120-NNs has the lowest reliability deviation, while the one utilizing 50-NNs possesses the overall best interval's pinball-loss. Figures~\ref{fig: reliability_deviation_interval_120NN} and \ref{fig: pinball_loss_Interval_120NN} show both evaluation values across all nominal coverages for the Poly1, ANN6, and SVR with $k_{NN}=120$, due to the fact that with such $k_{NN}$ the majority of techniques obtain their most accurate models. Poly1 and ANN6 are chosen additionally, because they possess the lowest reliability deviation for their type. Furthermore, Figures~\ref{fig: reliability_deviation_Interval_SVR} and \ref{fig: pinball_loss_Interval_SVR} plot the evaluation values against nominal coverage for SVR with varying amounts of nearest-neighbors.

\begin{figure}[!tb]
	\centering
	\includegraphics[width=0.8\textwidth]{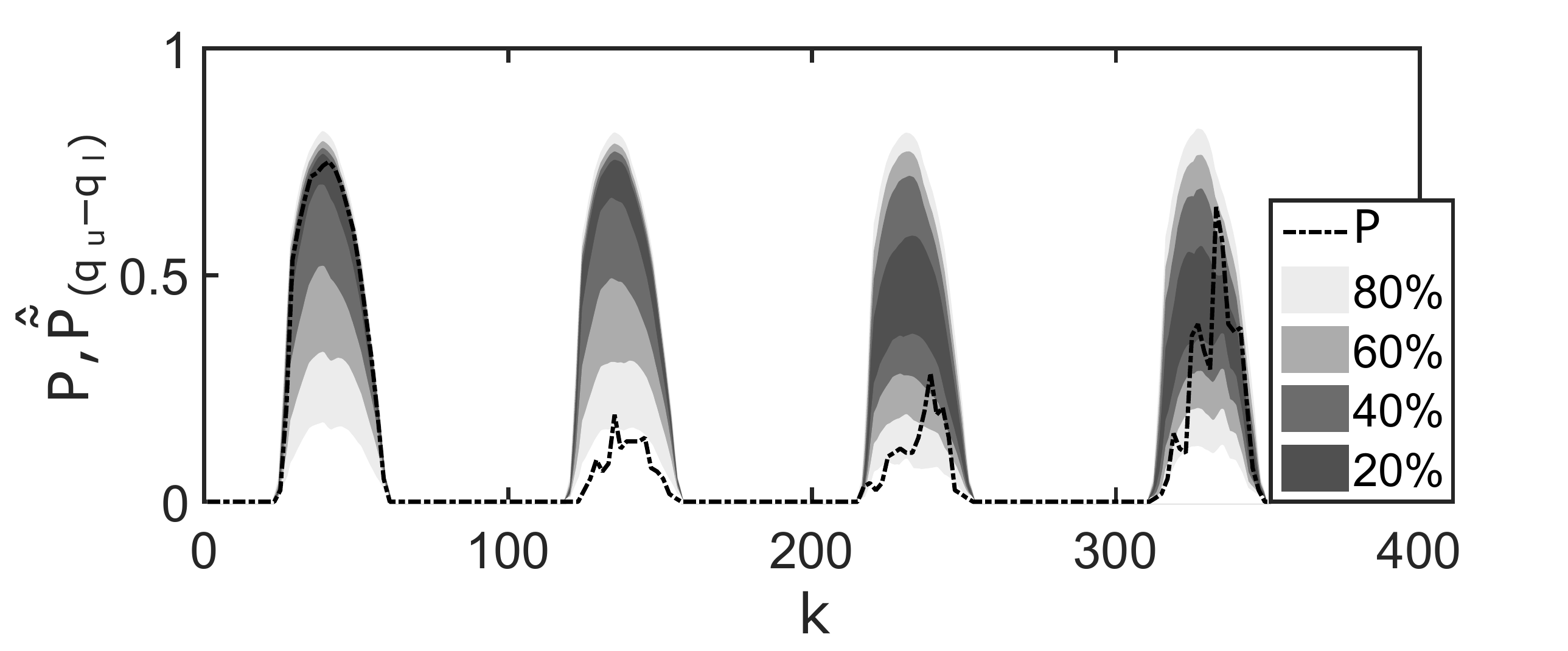}
	\caption{SVR 120-NNs interval forecasts for different nominal coverage ($0.2,0.4,0.6,$ and $0.8$)}
	\label{fig: forecast_example}
\end{figure}

Figures~\ref{fig: reliability_deviation_interval_120NN} and \ref{fig: pinball_loss_Interval_120NN} exemplify how the non-linear approaches are able to deliver more accurate interval forecasting models independently of the desired nominal coverage. Additionally, Figure~\ref{fig: reliability_deviation_interval_120NN} shows that techniques create underestimating intervals independently of the desired nominal coverage; a property which can be attributed to the higher reliability deviation of the lower quantiles in comparison to the higher ones. Furthermore, Figure~\ref{fig: reliability_deviation_Interval_SVR} depicts how the increasing number of nearest-neighbors appear to decrease the reliability deviations for a nominal coverage greater than approximately $0.3$, while Figure~\ref{fig: pinball_loss_Interval_SVR} shows an independence of the interval's pinball-loss (at least for the SVR case) towards the amounts of nearest-neighbors used in the present contribution. Moreover, the fact that the interval's pinball-loss for a nominal coverage higher than approximately $0.4$ was lower than the nominal coverage shows that trivial interval forecast are not created. The reason behind the high pinball-loss values for intervals with a nominal coverage lower than $0.4$ can be attributed to the great amount of values outside of them; by visualizing their results it can be concluded that non-trivial intervals are created. An example of obtained interval forecasts can be seen in Figure~\ref{fig: forecast_example}. It displays interval forecasts obtained with SVR and 120-NNs for a nominal coverage of $0.2,0.4,0.6,$ and $0.8$.

Even though the SVR technique seems to provide the most accurate quantile regressions and interval forecasting models, its long training times pose a major drawback. For example, the creation of the $99$ SVR quantile regressions without the forward feature selection using the present work's method takes approximately $24min$, when utilizing a Intel Core i7-4790 3.60GHz processor and 16Gb of RAM. In comparison, the training of the same $99$ quantile regressions with the forward feature selection takes approximately $1min$ for the polynomial techniques, while for the artificial neural networks it takes approximately $12min$.

Finally as a last remark, it is important to mention that by using the present contribution's methodology, the obtainment of quantile regressions with non-linear and complex data mining techniques (i.e. ANN and SVR) is simplified and made possible without changing the traditional algorithms used for their model's training. 

\section{Conclusion and Outlook}

The present contribution offers a simple methodology for the obtainment of data-driven interval forecasting models by combining pairs of quantile regressions. Those regressions are created without the usage of the non-differentiable pinball-loss function, but through a k-nearest-neighbors based training set transformation and traditional regression approaches. By leaving the underlying training algorithms of the data mining techniques unchanged, the presented approach simplifies the creation of quantile regressions with more complex techniques (e.g. artificial neural networks). The quality of the presented methodology is tested on the usecase of photovoltaic power forecasting, for which quantile regressions using polynomial models as well as artificial neural networks and support vector regressions are created. From the resulting evaluation values it can be concluded that acceptable interval forecasting models are created.

It is important to mention, that all quantile regressions for a specific household, data mining technique, and number of nearest-neighbors, are created with the same selected features. For such reason, future works should examine how much does the quantile regressions' and interval forecasts' quality increases or decreases if each quantile regression is able to select its optimal features separately. Another aspect which requires further inquiry is that the interval forecasting models obtained from using the present contribution's method are only able to quantify the uncertainty of the model's output given some uncertainty-free input values. Therefore, research regarding the quantification and propagation of input values uncertainties (e.g. uncertainty in forecast weather data) and ways to differentiate the uncertainty coming from the input values and the one coming from the model itself has to be conducted further. Also, the testing of the presented methodology on a benchmark dataset has to be carried out in order to identify its shortcomings and develop further improvements. Additionally, future PV power related works should test the methodology with different forecast horizons as well as different inputs (e.g. forecast solar irradiation) for the purpose of testing its behavior and its quality.

%\footnotesize
\textbf{Acknowledgments:}~\\ The present contribution is supported by the Helmholtz Association under the  Joint Initiative ``Energy System 2050 - A Contribution of the Research Field Energy''. 
\normalsize

%% Bibliography

%\includePaperPDF{paper_pdf}{\AuthorsTOC}{\AffiliationsTOC}{\Title}

\end{document}